\documentclass[reprint,showkeys,prb,floatfix]{revtex4-1}
\usepackage{verbatim}
\usepackage{amsfonts}
\usepackage{amssymb}
\usepackage{amsmath}
\usepackage{graphicx}
\usepackage[pdftex, pdftitle={Article}, pdfauthor={Author}]{hyperref} 
\begin{document}
\title{A simple recipe to create three-dimensional reciprocal space maps}

\author{Rafaela F. S. Penacchio}
    \affiliation{University of S{\~a}o Paulo, Institute of Physics, S{\~a}o Paulo, SP, Brazil}

\author{Celso I. Fornari}
    \affiliation{Experimentelle Physik VII and W{\"u}rzburg Dresden Cluster of Excellence ct.qmat, Fakult{\"a}t f{\"u}r Physik und Astronomie, Universit{\"a}t W{\"u}rzburg, W{\"{u}}rzburg, Germany}

\author{Maurício B. Estradiote}
    \affiliation{University of S{\~a}o Paulo, Institute of Physics, S{\~a}o Paulo, SP, Brazil}

\author{Philipp Kagerer}
    \affiliation{Experimentelle Physik VII and W{\"u}rzburg Dresden Cluster of Excellence ct.qmat, Fakult{\"a}t f{\"u}r Physik und Astronomie, Universit{\"a}t W{\"u}rzburg, W{\"{u}}rzburg, Germany}

\author{Marco Dittmar}
    \affiliation{Experimentelle Physik VII and W{\"u}rzburg Dresden Cluster of Excellence ct.qmat, Fakult{\"a}t f{\"u}r Physik und Astronomie, Universit{\"a}t W{\"u}rzburg, W{\"{u}}rzburg, Germany}

\author{Simon M{\"u}ller}
    \affiliation{Experimentelle Physik VII and W{\"u}rzburg Dresden Cluster of Excellence ct.qmat, Fakult{\"a}t f{\"u}r Physik und Astronomie, Universit{\"a}t W{\"u}rzburg, W{\"{u}}rzburg, Germany}
 
\author{Friedrich Reinert}
    \affiliation{Experimentelle Physik VII and W{\"u}rzburg Dresden Cluster of Excellence ct.qmat, Fakult{\"a}t f{\"u}r Physik und Astronomie, Universit{\"a}t W{\"u}rzburg, W{\"{u}}rzburg, Germany}
    
\author{S{\'{e}}rgio L. Morelh{\~{a}}o}
    \email[Corresponding author: ]{morelhao@if.usp.br}
    \affiliation{University of S{\~a}o Paulo, Institute of Physics, S{\~a}o Paulo, SP, Brazil}

\date{\today} 

\begin{abstract}
Combinations of advanced X-ray sources and zero-noise detector of enormous dynamic range have significantly increased the opportunity of mapping the reciprocal space of crystal lattices. It is particularly important in the design of new devices based on epitaxial thin films. In this work, we present a simple approach to the three-dimensional geometry involved in plotting reciprocal space maps, along with an example of computer codes to allow X-ray users to write their own codes. Experimental data used to illustrate an application of the codes are from antimony telluride epitaxial film on barium difluoride substrate.
\end{abstract}

\keywords{Synchrotron X-ray diffraction, nanostructured devices, area detectors, epitaxial films, topological insulators}

\maketitle

\section{introduction}


Acquisition of X-ray diffraction data with area detectors has become extremely popular nowadays \cite{cs09,dk13}. However, data analysis in the three-dimensional (3D) reciprocal space can be quite challenging for new users of X-ray facilities doted with this capability. It is particularly relevant to study nanostructured devices made of different materials and crystalline domains \cite{sb15,sm17,cf20a,ag19}. In this very short work, we put together the most straightforward information that a beginner X-ray user needs to know for processing their diffraction data into 3D plots of the reciprocal space, that is, to create a 3D reciprocal space map (3D-RSM).

\section{RSM basic 3D geometry}

\begin{figure*}
    \includegraphics[width=.98\textwidth]{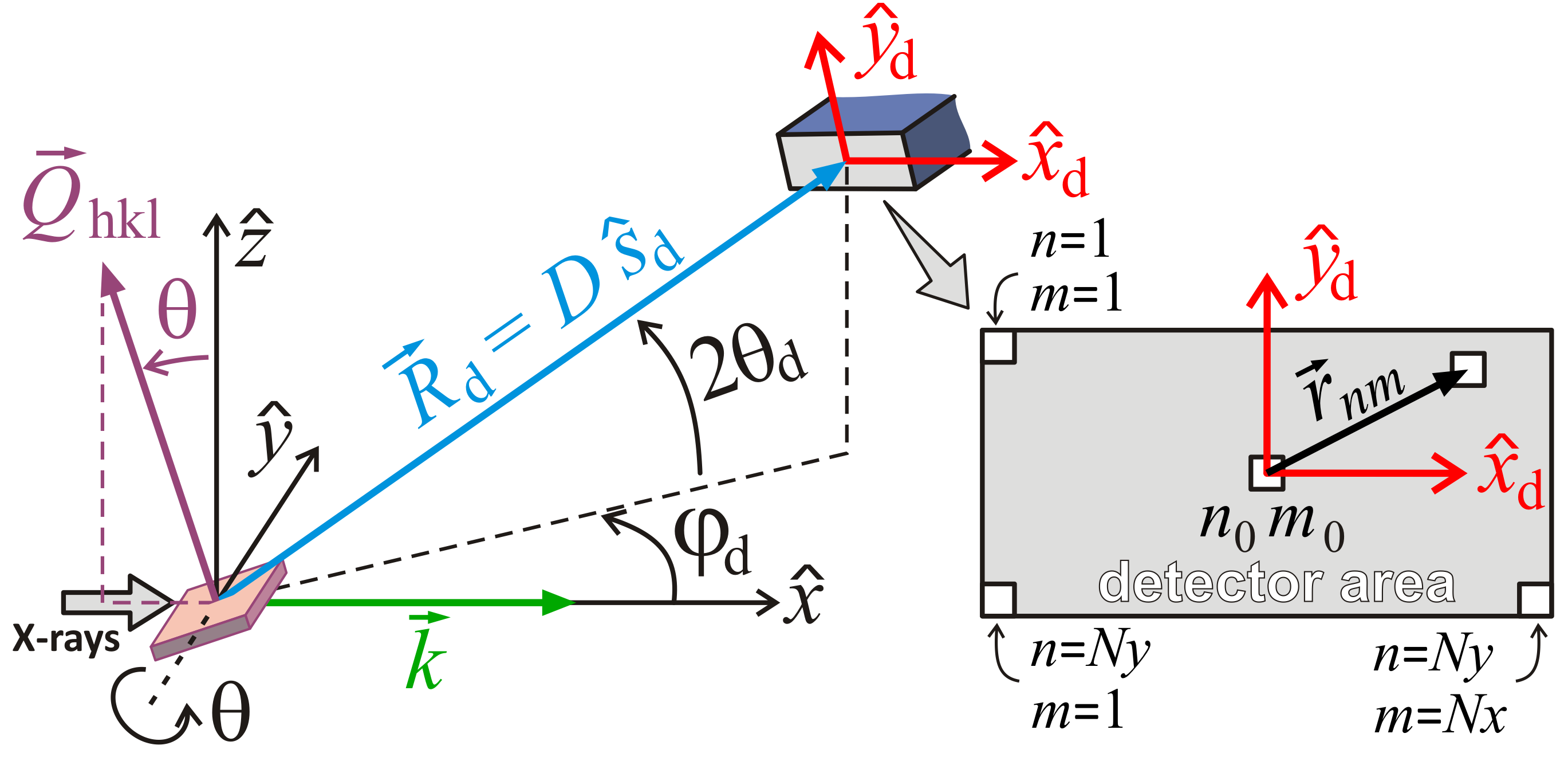}
    \caption{Lab reference frame $\hat{x}$, $\hat{y}$, and $\hat{z}$. Incident wavevector $\vec{k}=(2\pi/\lambda)\hat{x}$ (along axis $x$) for X-ray of wavelength $\lambda$. Spatial position of pixel $nm$ at the $N_y \times N_x$ pixel array of the detector area given as $\vec{r}_{nm} = (n_0-n)p\hat{y}_d + (m-m_0)p\hat{x}_d$ where $p$ is the pixel size, and $n_0 m_0$ stands for the pixel hit by the direct beam when $2\theta_d=0$ and $\varphi_d=0$.}
    \label{fig:labframe}
\end{figure*}

As schematized in Fig.~\ref{fig:labframe}, the spatial position of pixel $n_0 m_0$ in the lab reference frame is $\vec{R}_d$, and of any other pixel given as $\vec{R}_{nm} = \vec{R}_d + \vec{r}_{nm}$. In terms of the sample-detector distance $D$, 
\begin{equation}\label{eq:Rnm}
    \vec{R}_{nm} = D\hat{s}_d - (n-n_0)p\hat{y}_d + (m-m_0)p\hat{x}_d
\end{equation}
where the unit vectors are written as a function of the instrumental angles $2\theta_d$ and $\varphi_d$, as also depicted in Fig.~\ref{fig:labframe}.
{\small
\begin{eqnarray}\label{eq:versors}
    \hat{s}_d & = & \cos(2\theta_d)[\cos(\varphi_d)\hat{x} + \sin(\varphi_d)\hat{y}] + \sin(2\theta_d)\hat{z}\,, \nonumber\\
    \hat{x}_d & = & \sin(\varphi_d)\hat{x} - \cos(\varphi_d)\hat{y}\,,\quad{\rm and} \\
    \hat{y}_d & = & -\sin(2\theta_d)[\cos(\varphi_d)\hat{x} + \sin(\varphi_d)\hat{y}] + \cos(2\theta_d)\hat{z}\,.\nonumber
\end{eqnarray}}

The directions from the origin of the lab reference frame where the X-rays hit the sample, $[X,Y,Z] = [0,0,0]$, to the matrix of pixels define the set of scattering vectors 

\begin{equation}\label{eq:kprime}
    \vec{k}_{n m} = \frac{2\pi}{\lambda}\frac{\vec{R}_{n m}}{|\vec{R}_{n m}|}\,,
\end{equation}
as well as the accessible $Q$-space
\begin{equation}\label{eq:Qnm}
        \vec{Q}_{n m}  =  \vec{k}_{nm}-\vec{k} = \frac{2\pi}{\lambda}\left(\frac{\vec{R}_{nm}}{|\vec{R}_{nm}|} - \hat{x}\right)
\end{equation}
for the detector at a ($2\theta_d$,$\varphi_d$) position  \cite{sm16}. In the lab frame, this accessible $Q$-space is just a section of the Ewald sphere, and it remains constant as long as the direct beam direction, X-ray energy, and detector position are kept fixed during RSM data acquisition. On the other hand, the accessible $Q$-space in the crystal's frame of reference has a volume that depends on how the crystal's rotation is performed for data acquisition. 

\begin{figure}
     \includegraphics[width=.4\textwidth]{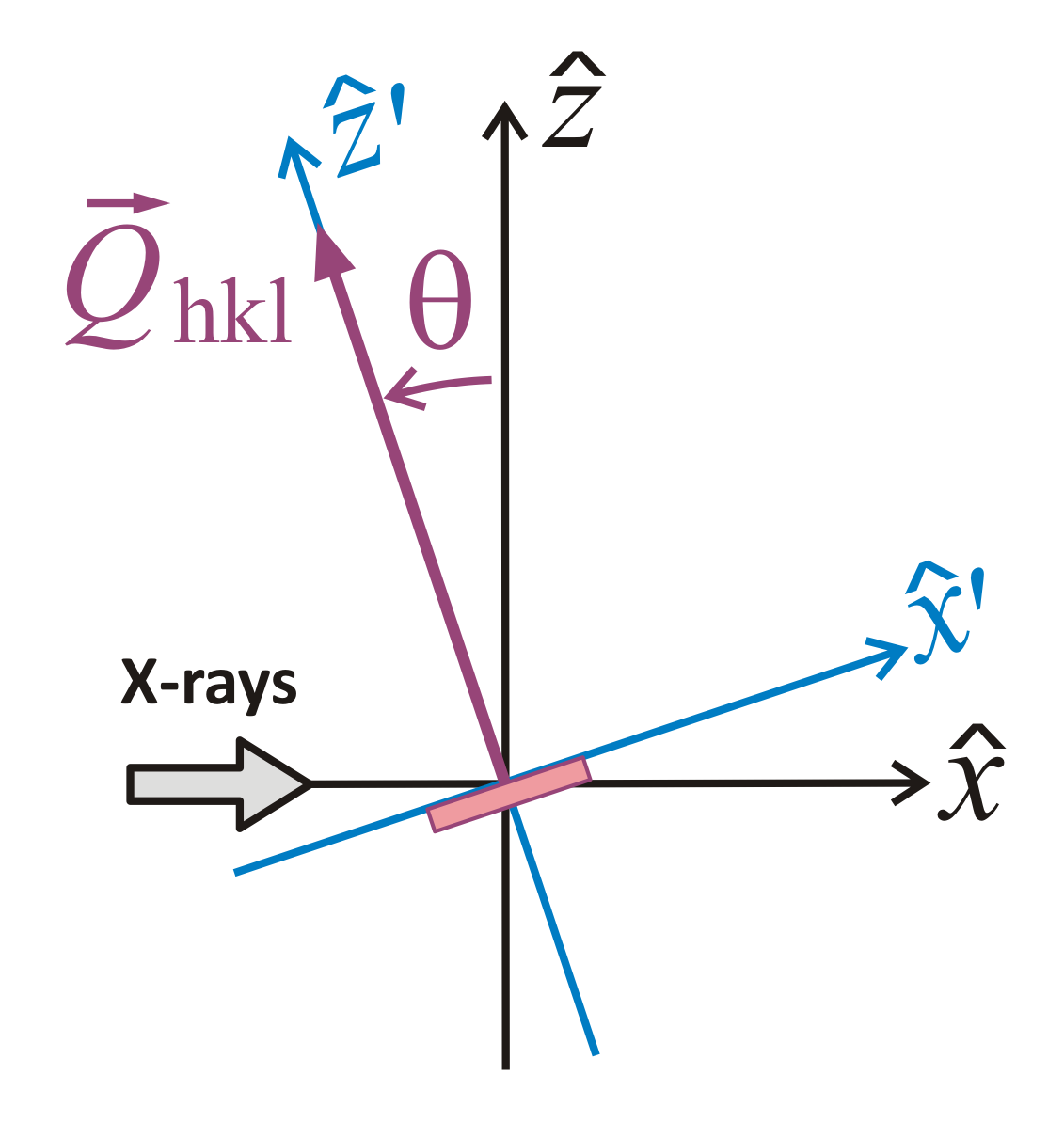}
    \caption{Crystal reference frame $\hat{x}^\prime$, $\hat{y}^\prime$, and $\hat{z}^\prime$ that undergoes a $\theta$ rotation around $y$ axis, $\hat{y}^\prime = \hat{y}$. Diffraction vector $\vec{Q}_{hkl} = (2\pi/d_{hkl})\hat{z}^\prime$ where $d_{hkl}$ is the atomic interplane distance of reflection $hkl$.}
    \label{fig:crystframe}
\end{figure}

In a simple rocking curve of $hkl$ reflection, the 3D-RSM technique aims to map the intensity distribution around a selected $hkl$ node. Its diffraction vector 
\begin{equation}\label{eq:Qhkl}
    \vec{Q}_{hkl} = Q_{hkl}\left[-\sin(\theta)\hat{x} + \cos(\theta)\hat{z}\right]
\end{equation}
is coplanar with the $xz$ plane of incidence, and it rotates around the $y$ axis with angle $\theta$ as defined in Fig~\ref{fig:crystframe}. The accessible $Q$-space in Eq~(\ref{eq:Qnm}) is projected in the crystal's frame $\hat{x}^\prime$, $\hat{y}^\prime$, and $\hat{z}^\prime$, as follow

\begin{eqnarray}\label{eq:DQxyz1}
    \Delta Q_x(n,m) & = & \vec{Q}_{n m} \cdot \hat{x}^\prime \nonumber \\
    \Delta Q_y(n,m) & = & \vec{Q}_{n m} \cdot \hat{y}^\prime \\
    \Delta Q_z(n,m) & = & \vec{Q}_{n m} \cdot \hat{z}^\prime\,. \nonumber 
\end{eqnarray}

\noindent To perform the dot products, these vectors have to be written in a common frame, that is 
\begin{eqnarray}\label{eq:xpypzp}
    \hat{x}^\prime & = & \cos(\theta)\hat{x}+\sin(\theta)\hat{z} \nonumber \\
    \hat{y}^\prime & = & \hat{y} \\
    \hat{z}^\prime & = & -\sin(\theta)\hat{x}+\cos(\theta)\hat{z} \nonumber 
\end{eqnarray}
resulting in
{\small
\begin{eqnarray}\label{eq:DQxyz2}
    \Delta Q_x(n,m)\! &\! =\! &\! (\vec{Q}_{n m} \cdot \hat{x})\cos(\theta) +  (\vec{Q}_{n m} \cdot \hat{z})\sin(\theta) \nonumber \\
    \Delta Q_y(n,m)\! &\! =\! &\! \vec{Q}_{n m} \cdot \hat{y} \\
    \Delta Q_z(n,m)\! &\! =\! &\! (\vec{Q}_{n m} \cdot \hat{z})\cos(\theta) - (\vec{Q}_{n m} \cdot \hat{x})\sin(\theta) \nonumber 
\end{eqnarray} 
}
as three matrices of reciprocal space coordinates of the pixels for each $\theta$ position of the rocking curve to which there is one matrix of intensity $I_{nm}$ from the detector image. To generate the 3D-RSM, it is necessary to compute the $\vec{Q}_{n m} \cdot \hat{\alpha}$  dot products ($\alpha = x,y,z$). According to Eq.~(\ref{eq:Qnm}), these products are constant values as a function of the $\theta$ rotation: 
\begin{eqnarray}\label{eq:Qxyz}
    \vec{Q}_{n m} \cdot \hat{x} & = & \frac{2\pi}{\lambda}\left(\frac{\vec{R}_{nm}\cdot\hat{x}}{|\vec{R}_{nm}|} - 1\right)\,, \nonumber \\
    \vec{Q}_{n m} \cdot \hat{y} & = & \frac{2\pi}{\lambda}\left(\frac{\vec{R}_{nm}\cdot\hat{y}}{|\vec{R}_{nm}|}\right)\,,\quad{\rm and} \\
    \vec{Q}_{n m} \cdot \hat{z} & = & \frac{2\pi}{\lambda}\left(\frac{\vec{R}_{nm}\cdot\hat{z}}{|\vec{R}_{nm}|}\right)\,. \nonumber
\end{eqnarray}

\section{Computer Codes Recipe}

Basic input data for coplanar ($\varphi_d=0$) 3D-RSM plots: 
\begin{enumerate}
    \item Wavelength $\lambda$.
    \item Elevation of the detector arm: $2\theta_d$. 
    \item Sample-detector distance $D$.
    \item Pixel size $p$. 
    \item Number of pixels $N_y$ (index $n=1,2,\ldots N_y$, running from top to bottom) and $N_x$ (index $m=1,2,\ldots N_x$, running from left to right) along $\hat{y}_d$ and $\hat{x}_d$ directions, respectively, as defined in Eq.~(\ref{eq:Rnm}).
    \item Central pixel indexes $n_0 m_0$ determined by setting both detector angles to zero.\\
    \item 1D array {\bf th}(1:N) containing the $N$ experimental $\theta_j$ angles of the rocking curve where $j=1,2,\ldots,N$. 
    \item 2D arrays of detector images {\bf Ij}(1:Ny,1:Nx) with the intensity data linked to each $j$ step. 
\end{enumerate}

For the $\theta$ rotation axis as defined in Fig.~\ref{fig:crystframe}, the ideal detector position is $2\theta_d = 2{\rm arcsin}\left(\lambda/2d_{hkl}\right)$ and $\varphi_d=0$. Matrices to be created, common to every $j$ step, are:
\begin{enumerate}
    \item 1D arrays {\bf sd}(1:3), {\bf xd}(1:3), and {\bf yd}(1:3) as the unit vectors in Eq.~(\ref{eq:versors}). Positions 1, 2, and 3 as the $x$, $y$, and $z$ componentes.
    \item 2D arrays {\bf Rx}(1:Ny,1:Nx), {\bf Ry}(1:Ny,1:Nx), and {\bf Rz}(1:Ny,1:Nx) as the matrices of $xyz$ components of the pixel positions $\vec{R}_{nm}$ in real space as in Eq.~(\ref{eq:Rnm}), and a 2D array {\bf R}(1:Ny,1:Nx) as the matrix of their modulus, $|\vec{R}_{nm}|$.
    \item 2D arrays {\bf Qx}(1:Ny,1:Nx), {\bf Qy}(1:Ny,1:Nx), and {\bf Qz}(1:Ny,1:Nx) as the matrices of $xyz$ components of the pixel position $\vec{Q}_{nm}$ in reciprocal space as defined in  Eq.~(\ref{eq:Qxyz}).
\end{enumerate}
Matrices to be created subsequently as a function of the rocking curve angle $\theta_j$ are:
\begin{enumerate}
    \item 3D arrays {\bf DQx}(1:Ny,1:Nx,1:N),\\ {\bf DQy}(1:Ny,1:Nx,1:N), and {\bf DQz}(1:Ny,1:Nx,1:N), exactly as shown in Eq.~(\ref{eq:DQxyz2}), for instance\\
    {\small $\textbf{DQx}(:,:,j) = \textbf{Qx}(:,:)\cos(\textbf{th}(j)) + \textbf{Qz}(:,:)\sin(\textbf{th}(j))$}.
    \item 3D array {\bf A}(1:Ny,1:Nx,1:N) with all concatenated images from the detector, that is\\ $\textbf{A}(:,:,j) = \textbf{Ij}(:,:)$.
\end{enumerate}
    
When these four 3D arrays: {\bf DQx}, {\bf DQy}, {\bf DQz}, and {\bf A} are stored in the computer's memory, available MatLab or Python packages can be used to plot the RSM. Next section shows an example using MatLab. 

General scripts to create 3D-RSMs by using either MatLab or Python are available for download at \href{https://github.com/rafaela-felix/rsm}{\it GitHub: Reciprocal space maps generator}.

\newpage
\onecolumngrid
\section{MatLab Codes and Example}\label{sec:3}

\begin{figure}[t]
    \includegraphics[width=.8\textwidth]{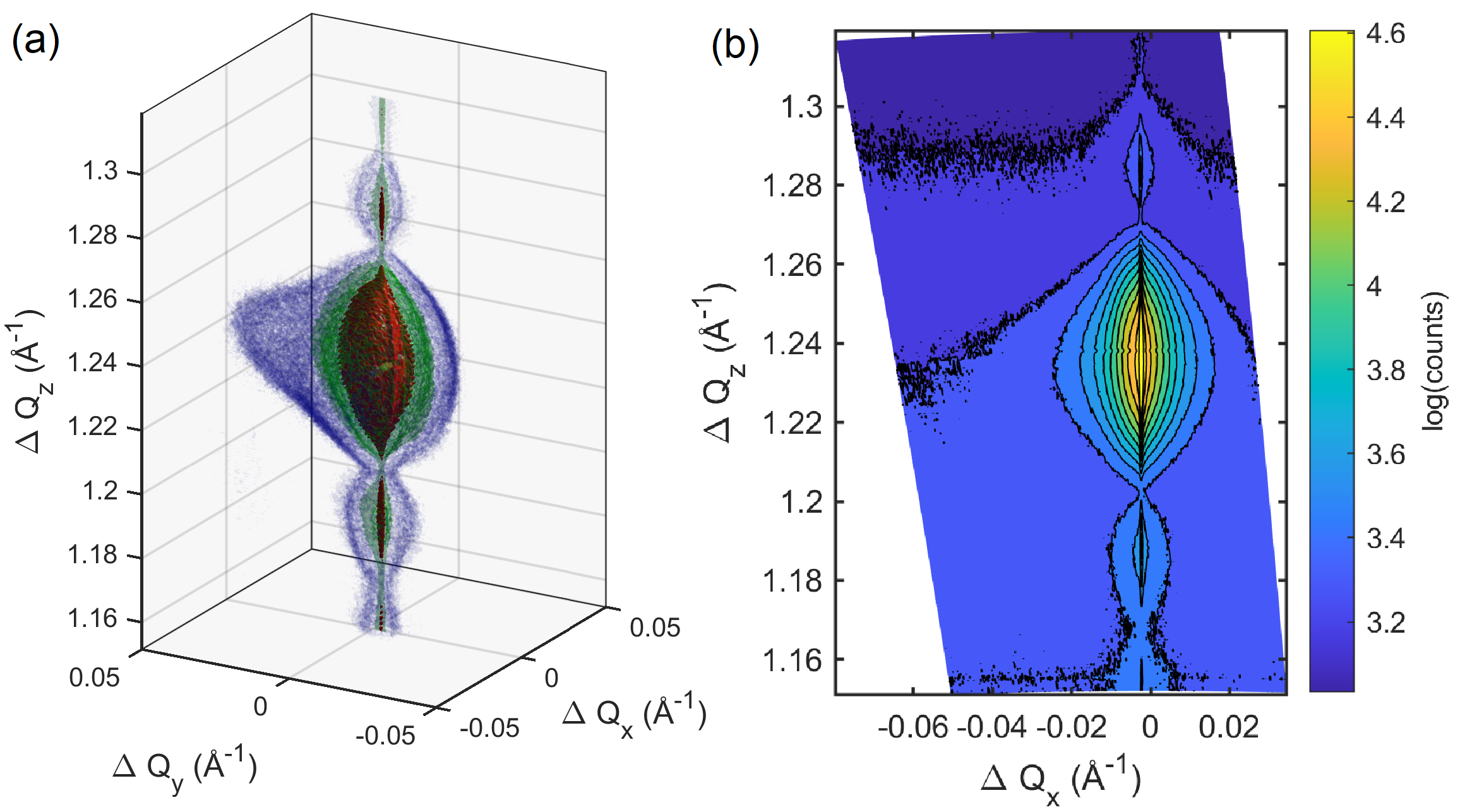}
    \caption{(a) 3D-RSM of 006 reflection of Sb$_2$Te$_3$ thin film on BaF$_2$ (111). (b) Projection of the RSM in the $xz$ plane of incidence. See \S \ref{sec:3} for details. DESY Photon Science, PetraIII beamline P08 (proposal I-20211588).  }
    \label{fig:3drsm}
\end{figure}

{\small 
\begin{verbatim}
function ReciprocalSpaceMaps
%%%%%%%%%%%%%%%%%%%%%%%%%%%%%%%%%%%%%%%%%%%%%
% Basic input info                          %
%%%%%%%%%%%%%%%%%%%%%%%%%%%%%%%%%%%%%%%%%%%%%
wl = 1.2401446;    % wavelength (Angstrom)
twoth_d = 14.037;  % elevation of detector arm (deg)
phi_d = 0;         % azimuth of detector arm (deg)
D = 1005.2; % sample-detector distance (mm)
p = 0.172;  % pixel size (mm)
Ny = 195;   % number of pixels along yd (vertical in Fig. 1)
Nx = 487;   % number of pixels along xd (horizontal in Fig. 1)
n0 = 95; m0 = 230; % central pixel (hit by the direct beam)
nb = 87; mb = 36;  % dead pixel 
folder='BiTe22040_00118\p100k'; % location of image files
fnamepng = 'BiTe22040L06';  % output file name
sclfactor = [.6 .45 .3];    % isointensity curve levels
Az = -60; El = 20;          % viewing angle of 3D plot
tifFiles = dir([folder '\*.tif']);
N = length(tifFiles);    % number of image files
dth = 0.01; % increment in the rocking curve angle th (deg)
th0 = 4.01852; % initial th value (deg)
th =  th0+(0:N-1)*dth;   % array of th values
%%%%%%%%%%%%%%%%%%%%%%%%%%%%%%%%%%%%%%%%%%%%%
% Common matrices to all th values          %
%%%%%%%%%%%%%%%%%%%%%%%%%%%%%%%%%%%%%%%%%%%%%
sd = [cosd(twoth_d)*[cosd(phi_d) sind(phi_d)] sind(twoth_d)];
xd = [sind(phi_d) -cosd(phi_d) 0];
yd = [-sind(twoth_d)*[cosd(phi_d) sind(phi_d)] cosd(twoth_d)];
ry = -([1:Ny] - n0)'*p;  % vertical pixels, 1st at the upper left corner
rx = ([1:Nx] - m0)*p;    % horizontal pixels
Rx = D*sd(1) + repmat(ry*yd(1),1,Nx) + repmat(rx*xd(1),Ny,1);
Ry = D*sd(2) + repmat(ry*yd(2),1,Nx) + repmat(rx*xd(2),Ny,1);
Rz = D*sd(3) + repmat(ry*yd(3),1,Nx) + repmat(rx*xd(3),Ny,1);
R = sqrt(Rx.*Rx + Ry.*Ry + Rz.*Rz);
aux = (2*pi/wl);
 Qx = aux*(Rx./R - 1);
 Qy = aux*Ry./R;
 Qz = aux*Rz./R;

%%%%%%%%%%%%%%%%%%%%%%%%%%%%%%%%%%%%%%%%%%%%%%%%%%%%%%%%%%%
% 3D matrices                                             %
%%%%%%%%%%%%%%%%%%%%%%%%%%%%%%%%%%%%%%%%%%%%%%%%%%%%%%%%%%%
DQx = zeros(Ny,Nx,N); DQy = DQx; DQz = DQx; A = DQx;
for j=1:N
  DQx(:,:,j) = Qx.*repmat(cosd(th(j)),Ny,Nx) + Qz.*repmat(sind(th(j)),Ny,Nx);  
  DQy(:,:,j) = Qy;
  DQz(:,:,j) = Qx.*repmat(-sind(th(j)),Ny,Nx) + Qz.*repmat(cosd(th(j)),Ny,Nx);
  Ij = double(imread([folder '\' tifFiles(j).name])); Ij(nb,mb) = 1;
  A(:,:,j) = Ij;
end

A(A<1)=1;
Amax=max(max(max(A)));
whalf = log10(0.5*Amax);
Z = log10(A);
hf1 = figure(1);
clf
set(hf1,'InvertHardcopy','off','Color','w')
patch(isosurface(DQx,DQy,DQz,Z,sclfactor(1)*whalf),'FaceColor','red',...
           'EdgeColor','none','FaceAlpha',1);
hold on
patch(isosurface(DQx,DQy,DQz,Z,sclfactor(2)*whalf),'FaceColor','green',...
           'EdgeColor','none','FaceAlpha',.22);
patch(isosurface(DQx,DQy,DQz,Z,sclfactor(3)*whalf),'FaceColor','blue',...
           'EdgeColor','none','FaceAlpha',.04);
hold off
       
view(3)
daspect([1 1 1])
axis tight
camup([0 0 1])
camlight
lighting phong
set(gca,'Color',[0.97 0.97 0.97],'Box','on',...
    'FontSize',10,'LineWidth',1,'FontName','Arial')
xlabel(['\Delta Q_x (' char(197) '^{-1})'],'FontSize',12)
ylabel(['\Delta Q_y (' char(197) '^{-1})'],'FontSize',12)
zlabel(['\Delta Q_z (' char(197) '^{-1})'],'FontSize',12)

camlight
axis tight
view(Az,El); 
grid; 
set(gca,'XLim',[-.05 .05],'YLim',[-.05 .05])
print('-dpng','-r300',[fnamepng '.png'])

%%%%%%%%%%%%%%%%%%%%%%%%%%%%%%%%%%%%%%%%%%%%%%%%%%%%%%%%%%%
% 2D projection                                           %
%%%%%%%%%%%%%%%%%%%%%%%%%%%%%%%%%%%%%%%%%%%%%%%%%%%%%%%%%%%
% Qz x Qx 
hf2 = figure(2); clf
set(hf2,'InvertHardcopy','off','Color','w')
XX=zeros(Ny,N);
ZZ=XX;
II=XX;
for n=1:N
    S = zeros(Ny,1);
    for m=1:Nx
        S = S + A(:,m,n);
    end
      II(:,n) = S; 
end
IImax = 0.5*max(max(II));
II(II>IImax) = IImax;
for n=1:N, XX(:,n) = DQx(:,m0,n); end
for n=1:N, ZZ(:,n) = DQz(:,m0,n); end
contourf(XX,ZZ,log10(II),12)
set(gca,'FontName','Arial','FontSize',10,'LineWidth',1);
axis image
axis tight
c = colorbar;
c.Label.String = 'log(counts)';
xlabel(['\Delta Q_x (' char(197) '^{-1})'],'FontSize',12)
ylabel(['\Delta Q_z (' char(197) '^{-1})'],'FontSize',12)
print('-dpng','-r300',[fnamepng 'QxQz.png'])


\end{verbatim}}



\bibliography{rsmmanuscript}






\end{document}